\documentclass[10pt]{revtex4}
\usepackage{graphicx}
\usepackage{amsmath,amssymb}
\usepackage{mathtools}

\begin{document}

\title{A Classical (Local) Account of The Aharonov-Bohm Effect}
\author{Vlatko Vedral}
\affiliation{Clarendon Laboratory, University of Oxford, Parks Road, Oxford OX1 3PU, United Kingdom\\Centre for Quantum Technologies, National University of Singapore, 3 Science Drive 2, Singapore 117543\\
Department of Physics, National University of Singapore, 2 Science Drive 3, Singapore 117542}

\begin{abstract}
It is frequently stated that the electromagnetic vector potential acquires a fundamental role in quantum physics, whereas classically it only represents a convenient, but by no means necessary, way of representing the electromagnetic field. Here we argue that this is a historical accident due to the fact that the electromagnetic field was discovered before photons, while the electron itself was discovered first as a particle, before it became clear that it must also be treated as a wave and therefore as an excitation of the underlying electron field.  We illustrate the fact that the vector potential ought to play a fundamental role classically using the Aharonov-Bohm effect. This effect is considered as the strongest argument for the role the vector potential plays in quantum physics, however, here we offer a fully classical account of it. This is a consequence of the fact that any account, be it classical or quantum, must involve the vector potential in order to preserve the local nature of the Aharonov-Bohm (as well as all the other) phases. 
\end{abstract}

\maketitle

Imagine that an alien civilisation discoverd physics much the same way as us, however, they had not yet discovered quantum physics, and the timelines for the discoveries of photons and electrons differed. Suppose that, like us, the aliens knew about the electromagnetic (EM) field (and wrote down what we call Maxwell's equations), but they didn't know about the existence of the individual fundamental discrete units of charge as sources of the electromagnetic field. They understood that an oscillating electrical current would generate an electromagnetic wave (as hypothesised by the alien Maxwell and demonstrated by the alien Hertz), however, they were unaware of the fact that the electrical current was made up of electrons (as were the earthlings Maxwell and Hertz, by the way). 

Crucially, however, the aliens comprehended that there must be an electron field. In other words, the alien de Broglie made the hypothesis that matter -- just like light -- is a wave before the alien Einstein introduced the idea of the quanta of light. Then, the alien G. P. Thompson experimentally confirmed the interference of electrial currents before his alien father J. J. discoved the electron (both received the alien Nobel Prizes but, unlike their earthly counterparts, first the alien son, followed by his alien father a few alian decades later).  
 
It might strike us as weird to think that the aliens introduced the electron field before they knew about the individual electrons, however, we humans did the same with the photon field (i.e. Maxwell's equations came before photons). This might be a simple accident due to the fact that, unlike electrons, photons are (loosely speaking) less interactive and have zero mass. But, it is possible to imagine that the alien technology, which is an extension of their senses (differing greatly from our human senses), finds it equally difficult to detect electrons. 

None of the reasons for the alien quirky physics history are really relevant for the present discussion other than the assumption that the alien physicists are treating both electrons and light as classial waves. They are, in fact, very pleased with themselves as they have arrived at a unified theory of everything in terms of waves (much to the astonishment of the alien Newton). While they have no inkling of the forthcoming quantum revolution (which ultimately also happens in the alien world), they already have their own alien Aharonov and their own alien Bohm. And the alien Aharonov and Bohm have discovered the (usual, not alien) Aharonov-Bohm effect \cite{AB}. They, with the help of the alien Chambers, diffracted the electron field of a crystal lattice behind which thay had inserted a solenoid whose magnetic field was confined within the solenoid (and strictly vanished outside). Just like our Aharonov and Bohm, the alien counterparts noticed that the electron field interference fringes were affected by the magnetic field of the solenoid even though the value of the magnetic field was zero in the region where the electron field was present. 

But this did not present a problem to them. The aliens could account for the Aharonov Bohm (AB) effect even though they fully believed in locality and knew no quantum physics. Namely the alien Einstein had already developed his theory of relativity and it was understood that no signals can travel faster than the speed of light. It was also clear to aliens that the field concept is crucial in implementing the notion of locality (i.e. avoiding any action at a distance, as emphasised by the alien Faraday). When fields interact they must do so only locally, i.e. point by point. Everything is a field, aliens reasoned, because relativity is a fundamental underlying symmetry in the universe. 

How does one account for the AB effect within the classical theory of fields? We simply write down the coupled field equations of the EM and the electron field and solve them. For our purposes, it will suffice to assume that the electron field is governed by the Klein-Gordon equation (the spin plays no role in the AB effect and so we do not need the full Dirac equation). It is normally thought that we need the complex electron field in order to couple to the EM field (because of the gauge invariance), however, there is a gauge in which the electron field is real (just like the electromagnetic one). We will call this the Schr\"odinger gauge as Schr\"odinger was the person to first introduce it \cite{Sch}. We want the electron field to be real simply because we would like it to be commensurate with the EM field (we do not want complex numbers to have to play any fundamental role). 

Let $A$ be the vector potential and $(\psi,\psi^*)$ the complex Klein-Gordon electron field. Then we define:
\begin{eqnarray}
\bar A_\mu (x) & = & A_\mu (x) + \frac{i}{2}(\ln \frac{\psi}{\psi^*})_{,\mu}\\
\phi (x) & = & |\psi (x)|
\end{eqnarray}
where the comma denotes the differentiation and $\phi$ is the real electron field. Note that the new vector potential is actually the old one ``dressed" with the electron field. It seems that the price of reality, given locality, is the loss of identity of the individual fields. 

We can now write the coupled equations in terms of the new fields:
\begin{eqnarray}
\Box \bar A_\mu (x) - \bar A^{\nu}_{,\mu\nu} & = & j_\mu(x) \\
(\Box +m^2) \phi (x) & = & e^2 \bar A^\mu (x) \bar A_{\mu}(x) \phi (x)
\end{eqnarray} 
where $j_{\mu}=-2e^2\bar A_{\mu} \phi^2$ is the current, $\Box$ represents the d'Alambertian and $m$ is the inverse Compton wavelength (deliberately not called the ``mass of the electron" as the aliens don't know about them). The interaction is of course still point-wise and the above equations are manifestly relativistically invarient. 

The key observation is that from the beginning we need to use the vector potential since this is the quantity that couples to the electron field in the point like manner, as in the term $A(x)\psi (x)$. We (and the aliens) know of no way of writing a point like interaction with the electric and the magnetic fields coupling to the fermionic field directly \cite{deWitt}. The discussion of the minimal coupling in Feynman \cite{Feynman} is instructive as it emphasises repeatedly that in quantum physics it is the potentials and not the forces that play the fundamental role. Feynman even says that he wished he had been taught the classical electrodynamics through potentials and not fields and forces, which is, unbeknownst to him, exactly how the alien Feynman was educated. 

The above equations can now be formally solved, more likely than not, perturbatively and to any desired order of accuracy. We remind the reader that all the quantities in the above equations are c-numbers, in other words all the component of each of the fields commute and can be measured simultaneously. This of course means that, while we can account for various phases such as the AB phase, we are unable to account for any proper quantum effects (to be discussed at the end). 

As far as the AB effect is concerned, let us assume without loss of generality that the electron field is confined to a narrow tube encircling the magnetic flux. Then all we need to do is compute the electron field and the $A$ field along the tube. The explanation as to how the AB phase is acquired is now identical to the quantum treatment \cite{ABMAVE} (see also \cite{Vedral-mach}). For us, it suffices to note that
when the electron field changes slowly (meaning that the amplitude change can be neglected and all that is left is the change in the phase of the electron wave), we can let $\Box \phi =0$ and the second of the field equations then reduces to
\begin{equation}
e^2 \bar A^\mu (x) \bar A_{\mu}(x) - m^2 =0 \; .
\end{equation}
The solution of this equation has a simple form: 
\begin{equation}
\psi(x)=e^{-ie \int_{x_0}^x A_\mu (x)dx^{\mu}} \psi_0 (x)\; ,
\end{equation}
which leads to the standard AB result when integrated over the closed loop starting at $x_0$ and ending at the electron field detection point (n.b. the real field $\phi$ contains trigonometric terms instead of the complex numbers, but the resulting interference would be just the same). The form is fully relativistic and includes the energy contribution, in this case simply given by the inverse Compton wavelength times Planck's constant and the speed of light (i.e. the rest mass of the electron times $c^2$, but as we said, there are no electrons here). It is also gauge invariant despite working in the particular Schr\"odinger gauge. Classical physics can therefore fully account for the AB phase providing we do not need to account for the individual electrons. 

The alians had an intuitive explanation of the AB effect too. This is because the alien Hamilton already realised that Newtonian mechanics can be phrased in terms of waves in which case various potentials can actually be seen as affecting the refractive index (the eikonal equation). Specifically, the difference between the refractive indices in regions with two different values of the vector potential, $A_1$ and $A_2$, is simply given by \cite{Kasunic}
\begin{equation}
\Delta n = n(A_1)-n(A_2) = \frac{p_0-eA_1}{p_0}-\frac{p_0-eA_2}{p_0}=\frac{e}{h}\Delta A\; ,
\end{equation}
where $p_0=h/\lambda_0$ is the electron's momentum when $A=0$. The AB shift is then
\begin{equation}
\Delta \theta = \frac{2\pi}{\lambda_0}t \Delta n 
\end{equation}
where $t$ is the ``thickness of the plate" (i.e. the region of interference). Of course, we do not need to talk about electrons at all; the change in the electron field wavelength will suffice. The expression for the AB phase is gauge invarient (without having to close the loop \cite{ABMAVE,Ehrenberg,Kasunic}) because it relies on the difference between vector potentials. This is of course always the case, since in any wave theory (and quantum waves inherit this too \cite{Vedral-mach}) only the relative phase has a physical meaning. The absolute phase, which would depend on the absolute value of the potential itself, is not measurable (neither classically nor quantumly). 

The refractive index in the electron wave theory being due to the change in the vector potential is in direct analogy to the optical refractive index being a function of the dielectric response. When electrons and light are treated on an equal footing then it is expected that each phenomenon present in the light field will have its analogue in the electron field. All this was, in fact, already understood perfectly well by Ehrenberg and Siday \cite{Ehrenberg}, who actually introduced the AB effect a decade before Aharonov and Bohm. 

Finally, a word of caution is in order. Given that the AB effect is considered a genuine quantum phenomenon, one might get the impression that our classical explination provides a local hidden variable model for quantum physics (see e.g. \cite{Akhmeteli}). Nothing could be further from the truth. All our theory needs to account for is the phase of the electron field acquired due to the interaction with the EM field. The phase itself is a classical concept and exists in any wave theory. Our classical theory here cannot account for the individual particles of either of the fields. This is because there are no q-numbers whose consequnce would be the non-commutativity of different field components, with particles being a manifestation of this complementarity. That, in turn, means that we are unable to account for the creation and annihilation of particles, nor can we account for the nature of their statistics (fermionic, bosonic). It also follows that the concept of entanglement cannot exist since entanglement between modes requires modes to contain complementary observables. Likewise, the zero-point energy is not there which results in the absence of the Casimir effect, the Lamb shift and the spontaneous emission of radiation. One might argue that the absence of the quantum vacuum is, in fact, the deadliest shortcoming of any classical field theory, since - in quantum field theory - all relevant quantities are ultimately constructed out of the vacuum to vacuum amplitudes for various products of the relevant field operators.

\textit{Acknowledgments}: The author thanks Chiara Marletto for many useful discussions on the topic of this paper. He also acknowledges funding from the National Research Foundation (Singapore), the Ministry of Education (Singapore) and Wolfson College, University of Oxford.

\end{document}